\documentclass{jnst} 
\usepackage[tablesfirst,notablist]{endfloat} 
\usepackage{bm}
\usepackage{ulem}
\usepackage{color}
\def\la{\mathrel{\mathpalette\fun <}}

\def\fun#1#2{\lower3.6pt\vbox{\baselineskip0pt\lineskip.9pt
\ialign{$\mathsurround=0pt#1\hfil##\hfil$\crcr#2\crcr\sim\crcr}}}
\def\dfrac#1#2{{\displaystyle\frac{#1}{#2}}}
\title{Deuteron-nucleus total reaction cross sections up to 1 GeV}
\author{Kosho Minomo$^{1}$\thanks{Corresponding author. Email: minomo@rcnp.osaka-u.ac.jp}, Kouhei Washiyama$^{2}$
and Kazuyuki Ogata$^{1}$}
\inst{$^{1}$Research Center for Nuclear Physics, Osaka University, Ibaraki 567-0047, Japan;\\
$^{2}$Center for Computational Sciences, University of Tsukuba, Tsukuba 305-8577, Japan
}
\abst{Total reaction cross sections of deuteron, $\sigma_d^{\rm R}$,
are calculated by a microscopic three-body reaction model.
The reaction model has no free adjustable
parameter and applicable to reactions at various
deuteron incident energies $E_d$ and with both stable and unstable nuclei.
The predicted $\sigma_d^{\rm R}$ are consistent with those evaluated by
a phenomenological optical potential for $E_d \la 200$~MeV in which
the potential has been parametrized.
A simple formula of $\sigma_d^{\rm R}$ up to $E_d = 1$~GeV,
as a function of $E_d$, the target mass number $A$ and its atomic number $Z$,
is given.}
\kword{deuteron; nuclear data; total reaction cross section; three-body reaction model; coupled-channel calculation; microscopic reaction theory}

\begin{document}
\maketitle
\newpage
\section{Introduction}
The total reaction cross section $\sigma^{\rm R}$ is one of the most
important physics observables for nuclear data studies. For proton-nucleus
reactions, $\sigma^{\rm R}$ has been investigated in many experiments found in the
literature (see, e.g., Refs.~\cite{Sih93,Car96,Mac99,Auc05,Abu10,Koh14} and references therein).
On the basis of about 1000 data values, Carlson~\cite{Car96} proposed a
simple formula of practical use for the proton-nucleus $\sigma^{\rm R}$,
$\sigma_p^{\rm R}$, at incident energies between 40~MeV and 560~MeV.
On the other hand, for deuteron-nucleus reactions, such systematic
studies on $\sigma^{\rm R}$ have not been performed. This is mainly
because of the lack of experimental data in comparison with those for
the proton-nucleus reactions. The global parametrizations of
the deuteron-nucleus optical potential by Daehnick {\it et al.}~\cite{Dae80}
and by Bojowald~{\it et al.}~\cite{Boj88} were highly successful but
limited up to around 90~MeV of the deuteron incident energy $E_d$.
Quite recently, a new global potential was developed by An and Cai~\cite{AC06}
that is applicable to the reactions up to $E_d=183$~MeV. However,
the maximum energy, i.e., 92~MeV per nucleon, is still much lower than that
of Carlson's formula for $\sigma_p^{\rm R}$. Furthermore, it should be
kept in mind that, in general, deuteron optical potentials are more ambiguous
than those for nucleon, since deuteron is a weakly-bound system and
the coupling to its breakup channels can affect the elastic scattering.
In other words, it is not trivial that one can describe the dynamical
polarization potential corresponding to the breakup channels
by a standard parametrization of the optical potential, e.g., the Woods-Saxon
form and its derivative.

To circumvent this, in the present study we propose to describe the
deuteron-nucleus reaction by means of a $p+n+{\rm A}$ three-body
reaction model, where A stands for the target nucleus. We adopt the
continuum-discretized coupled-channels method
(CDCC)~\cite{Kam86,Aus87,Yah12} that has successfully been applied to
deuteron-nucleus reactions in a wide range of energies. CDCC is a
non-perturbative reaction model that treats the couplings to projectile
breakup channels explicitly. The theoretical foundation of CDCC is given
in Refs.~\cite{Aus89,Aus96}, and the reaction observables calculated
by CDCC are shown~\cite{Del07} to agree with those by the Faddeev
theory~\cite{Fad61}, i.e., the exact solution of the three-body scattering
problem.
The input of the CDCC calculation is the nucleon-nucleus
($N$-A) optical potential, for which well-established global
potentials such as the parametrization by Koning and Delaroche~\cite{KD03}
and the so-called Dirac phenomenology~\cite{Coo09} are available.
In this study, however, we adopt a microscopic $N$-A potential
so as to calculate potentials for not only stable but also unstable
nuclei. This method is regarded as an application of the
microscopic reaction theory based on the nucleus-nucleus multiple
scattering theory~\cite{Yah08} to the deuteron-induced reactions.
Thus, we calculate unambiguously the deuteron-nucleus $\sigma^{\rm R}$,
$\sigma_d^{\rm R}$, for various target nuclei and for
$10 \le E_d \le 1000$~MeV. Then we parametrize the resulting
$\sigma_d^{\rm R}$ as a function of $E_d$, the target mass number $A$
and its atomic number $Z$.

In Section~2, we briefly describe the framework of the reaction model
and numerical inputs. We show in Section~3 typical results for
$\sigma_d^{\rm R}$ and give its functional form for practical use.
Section~4 is devoted to summary.

\section{Three-body description of deuteron-nucleus reactions}

In CDCC the total wave function $\Psi$ of
the $p+n+{\rm A}$ three-body system is expanded in terms of
the set $\{ \phi_i \}$ of the eigenstates of the internal Hamiltonian $h$
of the $p$-$n$ system:
\begin{equation}
\Psi({\bm r},{\bm R})
=\sum_{i=0}^{i_{\rm max}} \phi_i({\bm r}) \chi_i({\bm R}),
\label{eq1}
\end{equation}
where ${\bm R}$ is the coordinate of the center-of-mass of the $p$-$n$
system relative to the target nucleus A and ${\bm r}$ is that of
$p$ to $n$. The index $i$ specifies the $p$-$n$ eigenstate; $i=0$
corresponds to the deuteron ground state and $i>0$ to the
discretized-continuum states of the $p$-$n$ system.
The expansion coefficient denoted by $\chi_i$ describes the
scattering wave function between the $p$-$n$ system
in the $i$th state and A.
The three-body Schr\"{o}dinger equation to be solved is given by
\begin{equation}
\left[
T_{\bm R}+U_p+U_n+h-E
\right]
\Psi({\bm r},{\bm R})
=0,
\label{eq2}
\end{equation}
where $T_{\bm R}$ is the kinetic energy operator regarding ${\bm R}$
and $E$ is the total energy of the three-body system.
CDCC is the ordinary coupled-channel description of the three-body
reaction with the discretization of the $p$-$n$ continua; for more details
of CDCC as well as its theoretical foundation,
see Refs.~\cite{Kam86,Aus87,Yah12,Aus89,Aus96}.

In Equation~(\ref{eq2}), $U_p$ ($U_n$) is the $p$-A ($n$-A) scattering potential
consisting of nuclear and Coulomb parts.
For the nuclear part, we adopt the single folding model with the Melbourne nucleon-nucleon
$g$-matrix interaction~\cite{Amo00} and the one-body density $\rho$ of the nucleus A.
The nucleon-nucleus microscopic optical potentials, thus, constructed are
shown to reproduce the elastic scattering observables for various reaction
systems with no free adjustable parameters~\cite{Amo00,Min10,Toy13}.
In this study, $\rho$ is obtained by solving Hartree-Fock-Bogoliubov equations
in coordinate space with SLy4 Skyrme energy density functionals~\cite{Ben03}.
We use the computer code {\sc lenteur}~\cite{BenUPB}, which enforces
time-reversal and spherical symmetries.
For odd nuclei, the so-called filling approximation is adopted.

The model space of CDCC is as follows. We include $s$-, $p$- and $d$-waves
of $\phi$ calculated with the Ohmura potential~\cite{Ohm70}.
The $p$-$n$ continua are truncated at $k=1.0$~fm$^{-1}$, where $k$ is
the $p$-$n$ relative wave number, and the width of the momentum bin
is set to 0.1~fm$^{-1}$. The maximum value of $r$ ($R$) is
taken to be 100~fm (200~fm).
The Coulomb breakup effects are included in all cases, and the resulting $\sigma_d^{\rm R}$ converges
with this model space within 1\%.

We have calculated $\sigma_d^{\rm R}$ for $^{9}$Be, $^{12}$C, $^{16}$O, $^{28}$Si, $^{40}$Ca, $^{56}$Fe,
$^{58}$Ni, $^{79}$Se, $^{90}$Zr, $^{93}$Zr, $^{107}$Pd, $^{116}$Sn, $^{120}$Sn, $^{135}$Cs and $^{208}$Pb
at incident energies $E_d$ from 10~MeV to 1000~MeV. We emphasize that
we have not introduced any free adjustable parameters in the present
calculation.

\section{Results}

\begin{figure}
\includegraphics{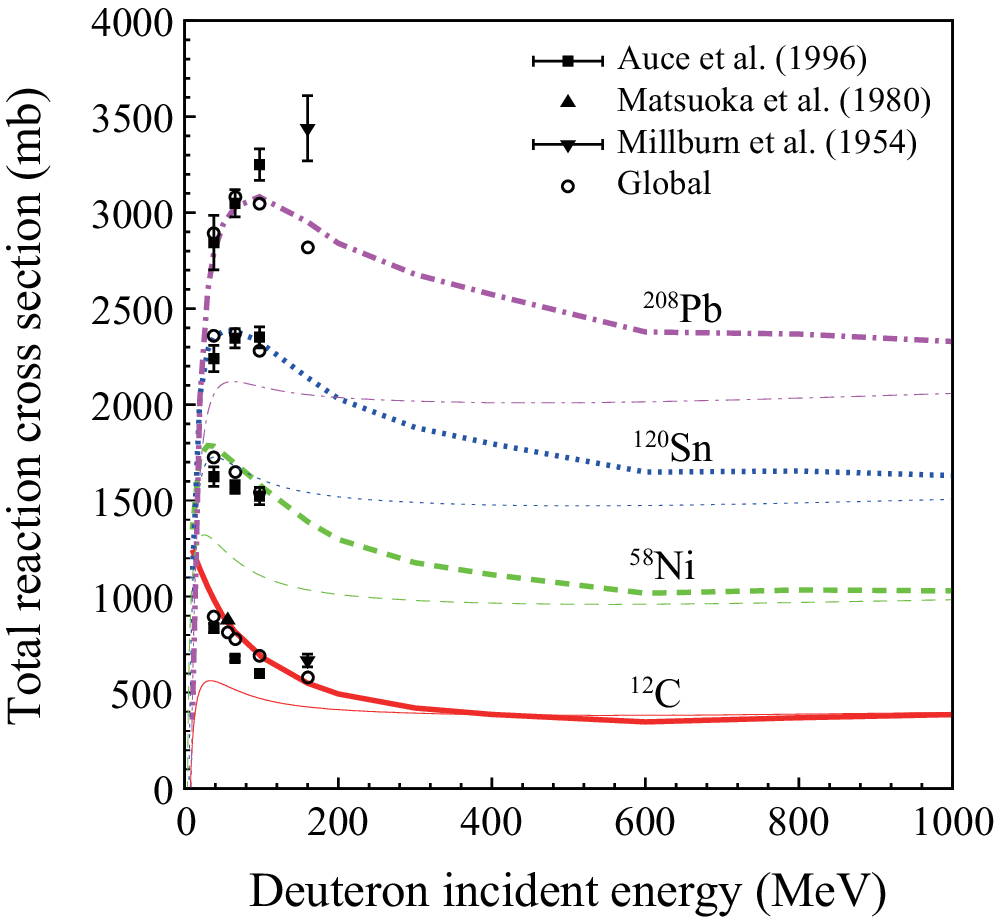}
\caption{
The predicted $\sigma_d^{\rm R}$ for
$^{12}$C (solid line), $^{58}$Ni (dashed line), $^{120}$Sn (dotted line) and $^{208}$Pb (dash-dotted line)
as a function of $E_d$.
The closed squares, triangles and inverted triangles are the experimental data
taken from Refs.~\cite{Auc96}, \cite{Mat80} and \cite{Mil54}, respectively.
The open circles represent the results calculated with the An-Cai global potential~\cite{AC06}.
The four thin lines represent the results of NASA's formula~\cite{Tri96,Tri97,Tri99}
implemented in PHITS~\cite{PHITS}.
}
\label{fig1}
\end{figure}

In Figure~\ref{fig1}, we show the predicted $\sigma_d^{\rm R}$ for
$^{12}$C, $^{58}$Ni, $^{120}$Sn and $^{208}$Pb by the
solid, dashed, dotted and dash-dotted lines, respectively,
as a function of $E_d$.
The experimental data taken from Auce {\it et al.}~\cite{Auc96}
(squares), Matsuoka {\it et al.}~\cite{Mat80} (triangles) and
Millburn {\it et al.}~\cite{Mil54} (inverted triangles) are shown by closed symbols.
The open circles represent the results calculated with the
An-Cai global potential~\cite{AC06}.
The prediction of the microscopic
CDCC calculation agrees well with the experimental data except for
the data measured at 160~MeV~\cite{Mil54}, at almost the same level
as that of the global optical potential~\cite{AC06}.
One sees from Figure~\ref{fig1}
that the $\sigma_d^{\rm R}$ for $^{12}$C measured at 160~MeV seems to deviate
from the energy dependence of the data at
lower energies, and that of the result of the An-Cai potential.
Systematic
measurement of $\sigma_d^{\rm R}$ at around 200~MeV with high precision
will be of great importance.
The four thin lines shown in Figure~\ref{fig1} represent the
results of NASA's formula~\cite{Tri96,Tri97,Tri99} implemented in the particle
and heavy ion transport code system (PHITS)~\cite{PHITS}; a severe
undershooting of the experimental data as well as the prediction of
the CDCC calculation is found.

\begin{figure}
\includegraphics{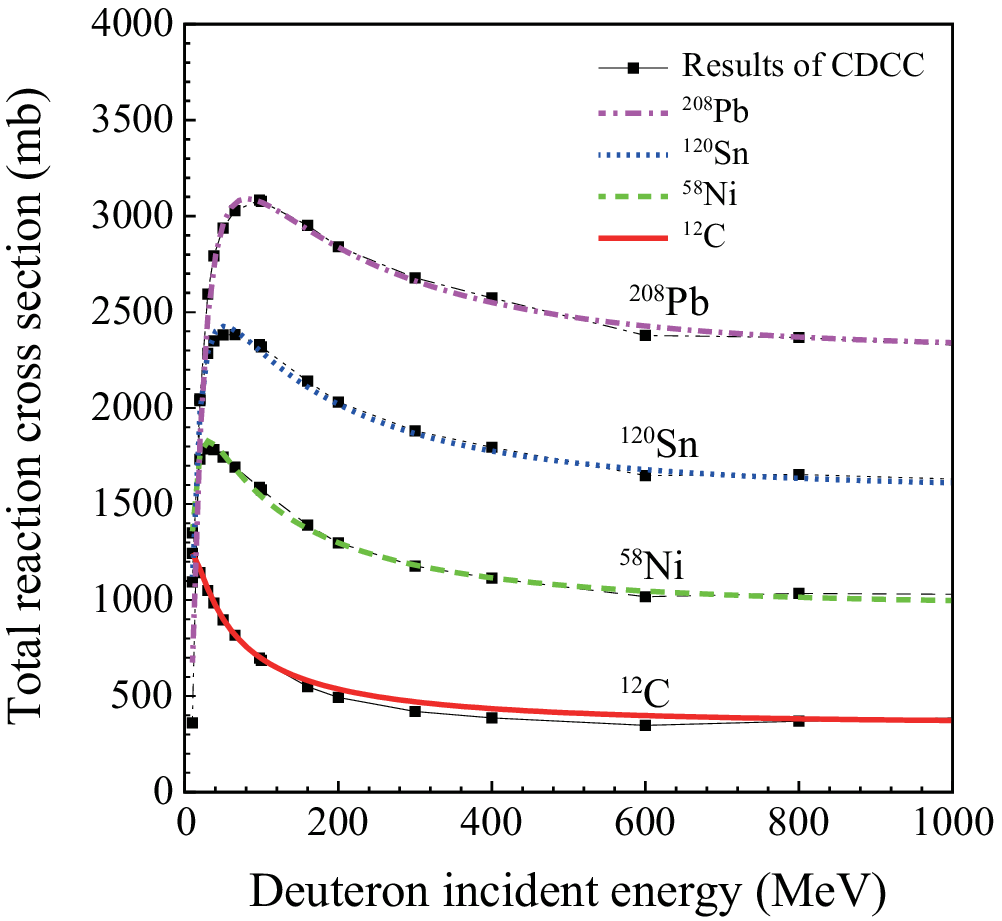}
\caption{
$\sigma_{\rm R}$ for
$^{12}$C (solid lines), $^{58}$Ni (dashed lines), $^{120}$Sn (dotted lines) and $^{208}$Pb (dash-dotted lines) targets,
as a function of $E_d$.
The thin and thick lines correspond to the
results of CDCC and those of Equation~(\ref{fit1}).
}
\label{fig2}
\end{figure}

Next we parametrize the $\sigma_d^{\rm R}$ calculated for
the 15 nuclei mentioned in Section~2 at
$10\le E_d \le 1000$~MeV by the following form:
\begin{equation}
\sigma_d^{\rm R}(E_d,A,Z)
=
\pi[R_d(E_d)+R_{\rm A}(E_d)A^{1/3}+\delta(E_d)A]^2C(E_d,Z),
\label{fit1}
\end{equation}
where $A$ ($Z$) is the mass number (atomic number) of the target nucleus A.
The effective radii of $d$ and A, and the correction term $\delta$ are defined by
\begin{equation}
R_d(E_d)=\dfrac{a_1}{1+a_2 \exp(-E_d/a_3)},
\end{equation}
\begin{equation}
R_{\rm A}(E_d)=\dfrac{b_1}{1+b_2 \exp(-E_d/b_3)}
\end{equation}
and
\begin{equation}
\delta(E_d)=\dfrac{c_1}{1+c_2 \exp(-E_d/c_3)},
\end{equation}
respectively.
The Coulomb damping factor is given by
\begin{equation}
C(E_d,Z)=\exp(-d_1 Z/E_d).
\label{fit2}
\end{equation}
This parametrization is quite similar to that of
Carlson's formula~\cite{Car96} for $\sigma_p^{\rm R}$ but
we have slightly changed the energy dependence of $R_d$ and $R_{\rm A}$,
and newly introduced $\delta$ and $C$. The ten parameters, $a_1$, $a_2$, $a_3$,
$b_1$, $b_2$, $b_3$, $c_1$, $c_2$, $c_3$ and $d_1$, determined by the least squares method,
are given in Table~\ref{tab1}.
The present formula well reproduces the
results of CDCC as shown in Figure~\ref{fig2}.
\begin{table}
\caption{Fitting parameters for $\sigma_d^{\rm R}(E_d,A,Z)$.}
\label{tab1}
\begin{center}
\begin{tabular}{c|ccc}
\hline
      & \multicolumn{3}{c}{$i$} \\
\cline{2-4}
      & 1          & 2        & 3       \\
\hline
$a_i$ & 0.306~fm   & $-0.923$ & 590~MeV \\
$b_i$ & 1.33~fm    & $-0.112$ & 248~MeV \\
$c_i$ & 0.00204~fm & $-0.788$ & 453~MeV \\
$d_i$ & 0.272~MeV  &          &         \\
\hline
\end{tabular}
\end{center}
\end{table}

\section{Summary}
We have calculated the deuteron-nucleus total reaction cross sections
$\sigma_d^{\rm R}$ for various target nuclei at deuteron incident energies
from 10~MeV to 1000~MeV, by means of the three-body reaction model, i.e.,
CDCC.
The nucleon-nucleus optical potential, which is the central input of CDCC
in the present study, was evaluated by the single-folding
model with the Melbourne $g$-matrix interaction and the nuclear one-body
density obtained by the Hartree-Fock-Bogoliubov method. Thus, the microscopic
description for the deuteron-nucleus reactions has been carried out.
The resulting values $\sigma_d^{\rm R}$ agree well with the experimental data
and the results of the An-Cai global optical potential.
NASA's formula for $\sigma_d^{\rm R}$ implemented in PHITS was found to severely
undershoot the results of CDCC, at low energies in particular.
We have parametrized our results of $\sigma_d^{\rm R}$ by a simple
functional form.

\section*{Acknowledgement}
The authors thank K. Niita, Y. Watanabe and K. Yoshida for useful discussions.
This work is supported in part by
Grant-in-Aid for Scientific Research (Nos. 16K05352 and 16K17698)
from the Japan Society for the Promotion of Science (JSPS)
and by the ImPACT Program of Council for Science,
Technology and Innovation (Cabinet Office, Government of Japan).
The numerical calculations in this work were performed at RCNP.

%


\begin{thebibliography}{99}

\bibitem{Sih93}
L. Sihver, C. H. Tsao, R. Silberberg, T. Kanai and F. Barghouty.
Total reaction and partial cross section calculations in proton-nucleus ($Z_t \le 26$) and nucleus-nucleus reactions ($Z_p$ and $Z_t \le 26$).
Phys. Rev. C 1993 March; 47: 1225-1236.

\bibitem{Car96}
R. F. Carlson.
PROTON-NUCLEUS TOTAL REACTION CROSS SECTIONS AND TOTAL CROSS SECTIONS UP TO 1 GeV.
At. Data Nucl. Data Tables 1996 May; 63: 93-116.

\bibitem{Mac99}
The GEM collaboration, H. Machner and B. Razen.
Absorption cross sections and efficiency of solid state detectors for light ions.
Nucl. Instru. Meth. A 1999 Nov; 437: 419-423.

\bibitem{Auc05}
A. Auce, A. Ingemarsson, R. Johansson, M. Lantz, G. Tibell, R. F. Carlson, M. J. Shachno, A. A. Cowley, G. C. Hillhouse, N. M. Jacobs, J. A. Stander, J. J. van Zyl, S. V. F\"{o}rtsch, J. J. Lawrie, F. D. Smit and G. F. Steyn.
Reaction cross sections for protons on $^{12}$C, $^{40}$Ca, $^{90}$Zr, and $^{208}$Pb at energies between 80 and 180 MeV.
Phys. Rev. C 2005 June; 71: 064606.

\bibitem{Abu10}
B. Abu-Ibrahim and A. Kohama.
Scaling properties of proton-nucleus total reaction cross sections.
Phys. Rev. C 2010 May; 81: 057601.

\bibitem{Koh14}
A. Kohama, K. Iida and K. Oyamatsu.
Energy and mass-number dependence of hadron-nucleus total reaction cross sections.
arXiv[nucl-th] 2014 Nov; 1411.7737.

\bibitem{Dae80}
W. W. Daehnick, J. D. Childs and Z. Vrcelj.
Global optical model potential for elastic deuteron scattering from 12 to 90 Mev.
Phys. Rev. C 1980 June; 21: 2253-2274.

\bibitem{Boj88}
J. Bojowald, H. Machner, H. Nann, W. Oelert, M. Rogge and P. Turek.
Elastic deuteron scattering and optical model parameters at energies up to 100 MeV.
Phys. Rev. C 1988 Sep; 38: 1153-1163.

\bibitem{AC06}
H. An and C. Cai.
Global deuteron optical model potential for the energy range up to 183 MeV.
Phys. Rev. C 2006 May; 73: 054605.

\bibitem{Kam86}
M. Kamimura, M. Yahiro, Y. Iseri, Y. Sakuragi, H. Kameyama and M. Kawai.
Chapter I Projectile Breakup Processes in Nuclear Reactions.
Prog. Theor. Phys. Suppl. 1986 Jan; 89: 1-10.

\bibitem{Aus87}
N. Austern, Y. Iseri, M. Kamimura, M. Kawai, G. Rawitcher and M. Yahiro.
Continuum-Discretized Coupled-Channels Calculations For Three-Body Models Of Deuteron-Nucleus Reactions.
Phys. Rep. 1987 Oct; 154: 125-204.

\bibitem{Yah12}
M.~Yahiro, K.~Ogata, T.~Matsumoto and K.~Minomo.
The continuum discretized coupled-channels method and its applications.
Prog. Theor. Exp. Phys. 2012 Sep; 2012: 01A206.

\bibitem{Aus89}
N. Austern, M. Yahiro and M. Kawai.
Continuum Discretized Coupled-Channels Method as a Truncation of a Connected-Kernel Formulation of Three-Body Problems.
Phys. Rev. Lett. 1989 Dec; 63: 2649-2652.

\bibitem{Aus96}
N. Austern, M. Kawai and M. Yahiro.
Three-body reaction theory in a model space.
Phys. Rev. C 1996 Jan; 53: 314-321.

\bibitem{Del07}
A. Deltuva, A. M. Moro, E. Cravo, F. M. Nunes and A. C. Fonseca.
Three-body description of direct nuclear reactions: Comparison with the
continuum discretized coupled channels method.
Phys. Rev. C 2007 Dec; 76: 064602.

\bibitem{Fad61}
L. D. Faddeev.
Scattering theory for a three particle system.
Zh. Eksp. Teor. Fiz. 1960 Nov; 39: 1459-1467.
[Sov. Phys. JETP 1961; 12: 1014-1019].

\bibitem{KD03}
A. J. Koning and J. P. Delaroche.
Local and global nucleon optical models from 1 keV to 200 MeV.
Nucl. Phys. A 2003 Jan; 713: 231-310.

\bibitem{Coo09}
E. D. Cooper, S. Hama and B. C. Clark.
Global Dirac optical potential from helium to lead.
Phys. Rev. C 2009 Sep; 80: 034605.

\bibitem{Yah08}
M.~Yahiro, K.~Minomo, K.~Ogata and M.~Kawai.
A New Glauber Theory Based on Multiple Scattering Theory.
Prog. Theor. Phys. 2008 Oct; 120: 767-783.

\bibitem{Amo00}
K.~Amos, P.~J.~Dortmans, H.~V.~Von Geramb, S.~Karataglidis and J.~Raynal.
Advances in Nuclear Physics, Vol. 25.
Plenum, New York: J.~W.~Negele and E.~Vogt; 2000. p.275-536.

\bibitem{Min10}
K.~Minomo, K.~Ogata, M.~Kohno, Y.~R.~Shimizu and M.~Yahiro.
Brieva-Rook localization of the microscopic nucleon-nucleus potential.
J. Phys. G 2010 July; 37: 085011.

\bibitem{Toy13}
M. Toyokawa, K. Minomo and M. Yahiro.
Mass-number and isotope dependence of local microscopic optical potentials for polarized proton scattering.
Phys. Rev. C 2013 Nov; 88: 054602.

\bibitem{Ben03}
M. Bender, P. H. Heenen and P. G. Reinhard.
Self-consistent mean-field models for nuclear structure.
Rev. Mod. Phys. 2003 Jan; 75: 121-180.

\bibitem{BenUPB}
K. Bennaceur (unpublished).

\bibitem{Ohm70}
T. Ohmura, B. Imanishi, M. Ichimura and M. Kawai.
Study of Deuteron Stripping Reaction by Coupled Channel Theory. II.
Prog. Theor. Phys. 1970 Feb; 43: 347-374.

\bibitem{Auc96}
A. Auce, R. F. Carlson, A. J. Cox, A. Ingemarsson, R. Johansson, P. U. Renberg, O. Sundberg and G. Tibell.
Reaction cross sections for 38, 65, and 97 MeV deuterons on targets from $^9$Be to $^{208}$Pb.
Phys. Rev. C 1996 June; 53: 2919-2925.

\bibitem{Mat80}
N. Matsuoka, M. Kondo, A. Shimizu, T. Saito, S. Nagamachi, H. Sakaguchi, A. Goto and F. Ohtani.
DEUTERON BREAK-UP IN THE FIELDS OF NUCLEI AT 56 MeV.
Nucl. Phys. A 1980 Aug; 345: 1-12.

\bibitem{Mil54}
G. P. Millburn, W. Birnbaum, W. E. Crandall and L. Schecter.
Nuclear Radii from Inelastic Cross-Section Measurements.
Phys. Rev. 1954 Sep; 95: 1268-1278.

\bibitem{Tri96}
R. K. Tripathi, F. A. Cucinotta and J. W. Wilson.
Accurate universal parameterization of absorption cross sections.
Nucl. Instru. Meth. B 1996 Oct; 117: 347-349.

\bibitem{Tri97}
R. K. Tripathi, J. W. Wilson and F. A. Cucinotta.
Accurate universal parameterization of absorption cross sections II - neutron absorption cross sections.
Nucl. Instru. Meth. B 1997 June; 129: 11-15.

\bibitem{Tri99}
R. K. Tripathi, F. A. Cucinotta and J. W. Wilson.
Accurate universal parameterization of absorption cross sections III - light systems.
Nucl. Instru. Meth. B 1999 Sep; 155: 349-356.

\bibitem{PHITS}
T. Sato, K. Niita, N. Matsuda, S. Hashimoto, Y. Iwamoto, S. Noda, T. Ogawa, H.
Iwase, H. Nakashima, T. Fukahori, K. Okumura, T. Kai, S. Chiba, T. Furuta, L.
Sihver, Particle and Heavy Ion Transport Code System PHITS, version 2.52, J.
Nucl. Sci. Technol. 2013 Sep; 50: 913-923.


\end{thebibliography}
\end{document}